\documentclass[twocolumn,smallcondensed,smallextended]{svjour3}
\smartqed  % flush right qed marks, e.g. at end of proof
\usepackage{graphicx,dcolumn,bm,color,latexsym,amssymb,amsmath}
\usepackage[FIGTOPCAP]{subfigure}
\usepackage[latin1]{inputenc}
\usepackage[english]{babel}

\definecolor{Blue}{rgb}{0.0,0.0,1}
\definecolor{Red}{rgb}{1,0.0,0.0}
\journalname{}

\begin{document}

\title{Normalization procedure for relaxation studies in NMR quantum information processing%\thanks{Grants or other notes
}


\author{A. Gavini-Viana         \and
        A. M. Souza         \and
        D. O. Soares-Pinto         \and
        J. Teles         \and
		R. S. Sarthour         \and
		E. R. deAzevedo         \and
		T. J. Bonagamba		\and
		I. S. Oliveira
}


\institute{A. Gavini-Viana \and R. S. Sarthour \and I. S. Oliveira\at
              Centro Brasileiro de Pesquisas F\'isicas, Rua Dr.
Xavier Sigaud 150, Rio de Janeiro 22290-180, RJ, Brazil\\
              %Tel.: \\
              %Fax: \\
              \email{gavini@cbpf.br}           %  \\
           \and
           A. M. Souza \at
				Institute for Quantum Computing and Department of
Physics and Astronomy,  University of Waterloo, Waterloo, Ontario,
N2L 3G1, Canada.
		   \and
		   D. O. Soares-Pinto \and E. R. deAzevedo \and T. J. Bonagamba   \at 
				Instituto de F\'isica de S\~ao Carlos, Universidade de
S\~ao Paulo, P.O. Box 369, S\~ao Carlos 13560-970, SP, Brazil
		   \and
		    J. Teles \at Instituto de Ciências Exatas e da Terra, Campus do Médio
 Araguaia, Universidade Federal de Mato Grosso, Rodovia MT-100, Pontal do Araguaia 78698-000, MT, Brazil	   
}			

\date{Received: date / Accepted: date}

\maketitle

\begin{abstract}
NMR quantum information processing studies rely on the reconstruction of the
density matrix representing the so-called pseudo-pure states (PPS).
An initially pure part of a PPS state undergoes unitary and
non-unitary (relaxation) transformations during a computation
process, causing a ``loss of purity'' until the equilibrium is
reached. Besides, upon relaxation, the nuclear polarization varies
in time, a fact which must be taken into account when comparing
 density matrices at different instants. Attempting to use
 time-fixed normalization procedures when relaxation is present,
 leads to various anomalies on matrices populations.
 On this paper we propose a method which takes into account the
time-dependence of the normalization factor. From a generic form for
the deviation density matrix an expression for the relaxing initial
pure state is deduced. The method is exemplified with an experiment
of relaxation of the concurrence of a pseudo-entangled state, which
exhibits the phenomenon of sudden death, and the relaxation of the
Wigner function of a pseudo-cat state.
\keywords{Pseudo-pure states \and NMR Relaxation
			\and Entanglement dynamics \and Decoherence}
\end{abstract}

%##########################################################################
\section{Introduction}
\label{intro}
The dynamics of open quantum systems  has been used for a long time
to study  physical ideas behind quantum measurements and
decoherence \cite{pazzurek01}. Special attention has been devoted to
those aspects of open quantum evolution which are relevant to
understand the transition from quantum to classical regimes.
Recently, the need for  the  maintenance of quantum state coherence,
for quantum computation  and communication tasks \cite{nielsen2000},
has increased  the interest in dynamical properties of open quantum
systems. In general, relaxation occurs as a result of the coupling of the
 system with the environment, which leads to loss of coherence, being
characterized mainly by energy relaxation and phase randomization
rates \cite{vandersypen04}. Energy dissipation carries away energy quanta on
 the scale of the Larmor frequency by enabling the spins to flip amongst 
their energy levels so as to
establish the equilibrium populations differences. Phase
randomization destroys the coherences due to inhomogeneities and
fluctuations in the environment. Furthermore, for nuclear spin systems with
$I>1/2$, additional quadrupolar interactions also leads
to relaxation when modulated by the environment. Recently,
theoretical as well as experimental studies about the nature of the
spin-environment coupling have been
reported \cite{zurek82,teklemariam03,nagele08}.

On the other hand, Nuclear Magnetic Resonance (NMR) has been
successfully used as an experimental method for many Quantum
Information Processing (QIP) implementations \cite{ivan2007}.
Generally, the NMR density matrix is written in the form:

\[ \rho_\epsilon = (1-\epsilon)\rho_I + \epsilon\rho_1 \]where
$\rho_I$ is the (normalized) identity matrix, and $\rho_1$ is an
\emph{arbitrary density matrix} \cite{braunstein99}. $\epsilon$ is a parameter varying
between 0 and 1, which is basically the ratio between the magnetic
and thermal energies. For liquid samples at room temperature,
$\epsilon \approx 10^{-5}$, what makes $\rho_\epsilon$ a highly
mixed state, close to the maximum entropy one.

In most NMR QIP implementations, the experiment is carried out in an
ensemble of spins in such a highly mixed state. Upon a set of
unitary and non-unitary operations, the density matrix
$\rho_1$ can be transformed into a pure state,
$|\psi\rangle\langle\psi|$. In this situation, $\rho_\epsilon$ is
called a \emph{pseudo-pure state} \cite{cory97}.

Pseudo-pure states allow exploring ensemble quantum computing using
NMR technique because, upon unitary transformations, which form the
basis of quantum logic gates and algorithms, only the pure deviation
matrix evolves. The identification of the pure part of the
pseudo-pure density matrix can be made from quantum state tomography
technique (QST).

However, PPS states are non-equilibrium states, and they relax back
to a Boltzmann distribution in a NMR characteristic time T$_1$, the
longitudinal (spin-lattice) relaxation time. Besides, NMR coherences vanish in a
time T$_2$, the transverse relaxation time. Regarding relaxation
studies of pseudo-pure states in the context of NMR quantum information processing, one can point out the studies reported on references
\cite{sarthour2003,ghosh2005,boulant03,auccaise2008}. In those
studies, relaxation behavior of various pseudo-pure states have been
investigated in different NMR systems. In reference
 \cite{sarthour2003} it was found  that spin-lattice relaxation
followed multi-exponential model that included mixed magnetic
dipolar and electric quadrupolar interactions. In Ref.
\cite{ghosh2005} it was found that whereas cross-correlations
accelerate the relaxation of certain pseudo-pure states, they delay
that of others. On the other hand, a robust method for quantum
process tomography provided a set of Lindblad operators that
optimally fit the density matrices measured at a sequence of time
points \cite{boulant03}. %Recently, Souza \textit{et al.} showed that
% the nuclear electric quadrupole relaxation can be interpreted as a
% computational process \cite{souza09}.

In spite of the various studies cited above, to the best of our
knowledge, there is no report showing how the initial pure part
evolves under relaxation processes. Yet, it is just this part of the
density matrix that matters for NMR QIP analysis. The reason is
that there is no simple normalization procedure to keep the
deviation part of $\rho_\epsilon$ a true density matrix along the
relaxation. Therefore, it would be desirable to have a way to follow
the evolution of an initial PPS state which undergoes a relaxation
process $|\psi\rangle\langle\psi|\rightarrow\rho_1$, yet keeping the
properties of a density matrix for the deviation, all the way
through. This would allow the study of other important quantum
information observables under relaxation. In this paper we propose
such a method. To illustrate it, we analyze experimental data for
the concurrence of initially pseudo-entangled states and the
discrete Wigner functions of a relaxing quadrupolar spin 3/2.

The paper is organized as follows: In Sec.\ref{secnorm} we show that
there is no trivial normalization procedure for the  deviation
density matrix, and that fixed normalization schemes
produces invalid density matrices. In Sec.\ref{secnormmet} we
develop a normalization procedure that produces valid relaxing
density matrices. Finally, in Sec.\ref{secexp} we present some application
examples of the method.

%###################################################################################
\section{Time-Independent Normalizing procedure for the deviation density matrix}
\label{secnorm}

The density matrix formalism is an appropriate description for both,
pure and mixed quantum states \cite{nielsen2000}. Here, we consider
a general density matrix for a spin ensemble with the form:

\begin{equation}
\rho=\rho_I + \sigma \label{eqdev}
\end{equation}

\noindent where $\rho_I=\mathbf{1}/N$ represents a maximally mixed
state and $\sigma$ is called \textit{deviation density matrix} \cite{suter2009}. This
type of density matrix appears, for instance, from spin ensemble
coupled to a heat reservoir at a fixed temperature and whose thermal
energy is much bigger than the magnetic energy of nuclear moments in
the ensemble. This is called the high-temperature approximation.

For some systems with zero trace observables, such as the case of
NMR, only $\sigma$ can be experimentally observed. This means that
quantum information quantities, such as the concurrence, entropy or
Wigner functions, among others,  cannot be directly obtained from
experimental data; for that a true density matrix is necessary.

We consider now a quantum process described by a trace preserving
quantum operation $\mathcal{S}$. Applying such operator to the
density matrix Eq.(\ref{eqdev}) yields

\begin{eqnarray}
\mathcal{S}\left(\rho\right)&=&\mathcal{S}\left(\rho_I\right) + \mathcal{S}\left(\sigma\right) \\
&=& \rho_I + \left[\mathcal{S}\left(\rho_I\right) - \rho_I +
\mathcal{S}\left(\sigma\right)\right],
\end{eqnarray}

\noindent which has the same form as (\ref{eqdev}), if we make the
identification:

\begin{equation}
\sigma \overset{\mathcal{S}}{\longrightarrow}
\left[\mathcal{S}\left(\rho_I\right) - \rho_I\right] +
\mathcal{S}\left(\sigma\right),
\end{equation}

\noindent This shows that the process $\mathcal{S}$ is not generally
linear in $\sigma$. The linearity of  $\mathcal{S}$ with respect to
$\sigma$ is only obtained if  $\mathcal{S}$ is a \emph{unital}
process ($\mathcal{S}(\mathbf{1})=\mathbf{1}$), for instance,
unitary transformations, which is not the case for most relaxation
phenomena \cite{garcia-mata2005}. Trace-preserving unital processes
are the quantum analogs of the classical doubly stochastic
processes, which can only increase the von-Neumann entropy of a
state (defined to be $-$Tr$\left\{\rho \text{log} \rho\right\}$).
Otherwise, non-unital processes provide the only means to reduce
entropy and prepare initial states \cite{leung}. Usually in quantum
algorithms applications, the identity term of the Eq.(\ref{eqdev}) can be neglected because
it is invariant under unital processes.

In order to illustrate a relaxation process $\mathcal{S}$ acting on
the density matrix, the deviation density matrix obtained from
Redfield theory for a purely quadrupolar relaxation of an ensemble
of spin $I=3/2$ will be considered (the deviation density matrix
elements are on Appendix \ref{aRed}). We consider as initial state a
pseudo-pure state

\begin{equation}
\rho_{\epsilon} = \left(1-\epsilon\right)\rho_I +
\epsilon|\psi\rangle\langle\psi|, \label{eqpps}
\end{equation}

\noindent where $|\psi\rangle$ is a pure state. It could be, for
instance, the pure state $|11\rangle$. From (\ref{eqpps}) we have:

\begin{eqnarray}
|11\rangle\langle 11| &=& \frac{1}{\epsilon}\left[
\rho_{\epsilon}-\left(1-\epsilon\right)\rho_I \right] \\
\label{eqppsdev}
                      &=& \frac{1}{\epsilon} \sigma + \rho_I .
\label{eqn1}
\end{eqnarray}

\begin{figure}[h!]
 \centering
 \includegraphics[scale=0.6]{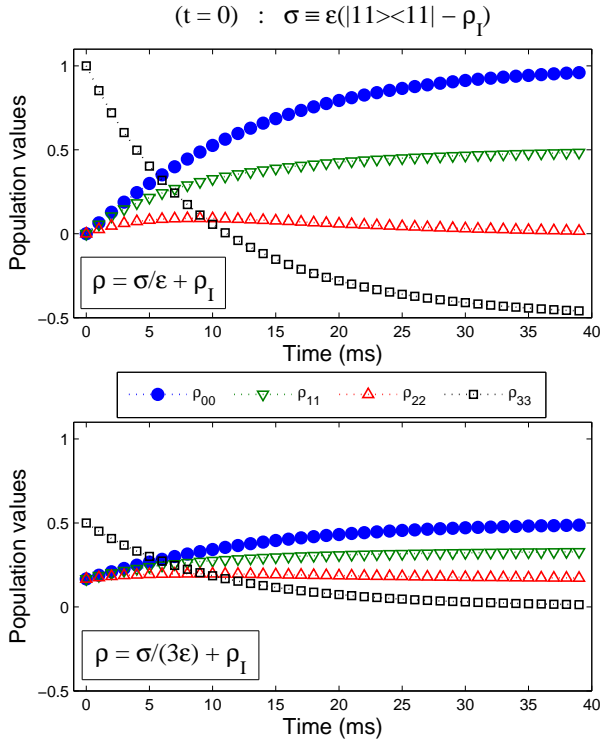} %,width=0.5\textwidth,height=0.25\textheight
 \caption{Calculated population values of the initial computational basis state $|11\rangle$ evolving under pure
 quadrupolar relaxation. The top and bottom plots represent two different types of time-fixed
  normalization to deviation density matrix, from which we obtain the pure part of the pseudo-pure state (see Appendix A
  for details). On the top scheme the matrix population become negative after 10 ms. On the bottom one populations are positive,
  but their initial values are incorrect. }
 \label{figpopeps}
\end{figure}

\noindent Hence, normalizing the deviation density matrix by
$\epsilon$ and adding to $\rho_I$ the pure state can be obtained.
But, since $\sigma$ changes with evolving time due to quadrupolar
relaxation process, the loss of purity will make the pure state
$|11\rangle$ to change to a mixed state. In Fig. \ref{figpopeps}
(top) is shown the relaxation of the population (diagonal matrix
elements) of the $|11\rangle$ state. This was obtained considering
fixed normalization, like Eq.(\ref{eqppsdev}) for $\sigma$ varying
in time. Since populations assume negative values at times above
$10$ ms, one no longer has a  valid density matrix. Therefore,
maintaining the normalization of the Eq.(\ref{eqppsdev}), leads to
matrices exhibiting negative populations at some instants of time. On
another hand, if we try adjusting the final population values, the
equilibrium state $\rho_{eq}$, we can use the following
normalization

\begin{equation}
\rho_{eq} = \frac{1}{\epsilon '}\sigma + \rho_I \label{eqn2}
\end{equation}

\noindent such that the $\rho_{eq}$ has trace equal $1$ and
non-negative populations. In this case (electric quadrupole
relaxation to $3/2$ nuclear spin)  $\epsilon ' = 3\epsilon$. It can
be seen in Fig. \ref{figpopeps} (bottom) that they have
non-negative values for the whole  interval of  time, but at $t=0$
we have incorrect populations values for the pseudo-pure state. Thus
we also discard this method of normalization.

From these two different ways to normalize  initial and final
deviation matrices,  we can conclude that the normalization must be
time-dependent.

%#############################################################################
\section{Time-Dependent Normalization Method}
\label{secnormmet}

First we will consider initial and final conditions for $\rho(t)$.
Re-writing (5) in a slightly different way,  we have the pseudo-pure
state,

\begin{eqnarray} \label{eqsig1}
\rho(t=0)&=&\rho_I +
\epsilon\left(|\psi\rangle\langle\psi|-\rho_I\right).
\end{eqnarray}

\noindent In words, the pseudo-pure state is composed by a maximum
statistical mixture, $\rho_I$, and  a pure contribution
$|\psi\rangle$.

Under relaxation, the system reaches the equilibrium state. For this
state one can consider the Zeeman interaction as the most important
one. In the high-temperature limit:

\begin{eqnarray} \label{eqsig2}
\rho(t \rightarrow \infty)&=&\frac{e^{\beta\hbar\omega
I_z}}{Tr\left(e^{\beta\hbar\omega I_z}\right)} \approx \rho_I +
\frac{\hbar\beta\omega}{N}I_z \nonumber  \\ \label{eqsig3}
                                 &=& \rho_I + 2I\epsilon\left(\rho_z - \rho_I\right),
\end{eqnarray}

\noindent where  $ I $ is the angular momentum of the nuclear spin
and $\epsilon=\hbar\beta\omega/2$, and $\rho_z \equiv (1/NI)I_z +
\rho_I$.

From this, we can write for the respective  initial and final
(equilibrium) deviation matrices: $\sigma_{0} =
\epsilon\left(|\psi\rangle\langle\psi|-\rho_I\right)$ and
$\sigma_{\infty} = 2I\epsilon\left(\rho_z - \rho_I\right)$. The
initial deviation matrix has ``polarization'' $\epsilon$, whereas
the equilibrium has ``polarization'' $2I\epsilon$. The relaxation
phenomenon connects these two matrices in time. Therefore, we can
write a general form for a deviation density matrix:

\begin{equation}\label{eqsig4}
\sigma= \alpha \left(\rho_{\alpha} - \rho_I\right),
\end{equation}

\noindent where $\alpha$ is the polarization and $\rho_{\alpha}$ the
respective density matrix. From $\alpha$ and  $\rho_I$, $\rho_{\alpha}$ can
be determined for any deviation density matrix $\sigma$.

Adding $\sigma_0$ to both sides of the Eq. (\ref{eqsig4}) and isolating $\rho_{\alpha}$, we
have:

\begin{equation}
\rho_{\alpha} = \frac{\epsilon}{\alpha}|\psi\rangle\langle\psi| +
\frac{1}{\alpha}\left[\sigma-\sigma_0\right] + \left(1 -
\frac{\epsilon}{\alpha}\right)\rho_I. \label{eqrhom}
\end{equation}

\noindent This is the final expression for the density
matrix for a given $\alpha$ and $\sigma$. For example, at $t=0$,
 $\alpha =\epsilon$ and $\sigma = \sigma_0$, and we
are left with a pure state. Upon relaxation the other terms come
into action. Notice that the first and last terms on the right side
can be interpreted as a depolarization channel \cite{nielsen2000},
whereas the middle term depends on the relaxation model, for
instance, the Redfield theory. $\alpha$ is a time-adjustable
parameter, which varies in the interval $(0, 2I\epsilon]$. It is
worth mentioning that there is a possible connection between
$\alpha$ and the effective number of nuclei per state
\cite{cory97}.

\begin{figure}
 \centering
 \includegraphics[scale=0.65]{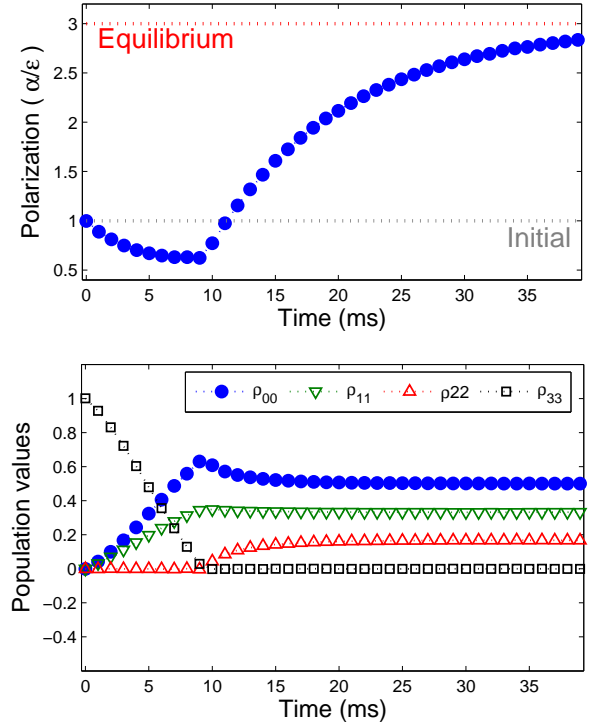}
 \caption{Time evolution of the polarization and respective normalized populations for spin $3/2$ nuclei
 in the initial state $|11\rangle\langle 11|$.
}
 \label{figpop11_polx}
\end{figure}

Eq.(\ref{eqrhom}) can be used to obtain the properly normalized
density matrix $\rho_{\alpha}$, at any instant of time, from the
measured deviation density matrices. A numerical example is shown on
Fig. \ref{figpop11_polx}, for the same parameters used to calculate
the data on Fig. \ref{figpopeps}. At any instant of time the
populations add $1$ and never become negative. Next we will analyze
experimental results for pseudo-entangled states and Wigner
functions.

%##########################################################################
\section{Experimental results}
\label{secexp}

$^{23}$Na NMR experiments were performed using a magnetic field
oriented lyotropic liquid-crystal system (Sodium Dodecyl Sulfate,
SDS), using a 9.4 T - VARIAN INOVA spectrometer. The sample
composition was 21.3$\%$ of SDS, 3.6$\%$ of decanol, and 75.1$\%$ of
deuterium oxide. The quadrupolar coupling was found to be (16700
$\pm$ 70) Hz at 24 $^{\circ}$C. In the experiments that
characterized the relaxation of all elements of the density matrix,
an initial pseudo-pure state was prepared using the SMP technique
 \cite{fortunato2002}. In this case, all the deviation density matrix
elements were measured using the QST method via coherence
selection \cite{teles2007}. The basic experimental scheme consisted
of: i) a state preparation period performed with SMP; ii) a variable
evolution period $\tau$ where relaxation takes place; and iii) a hard RF
pulse with the correct phase cycling and duration to execute QST via
coherence selection, as described in detail in Ref.
 \cite{auccaise2008}.

\subsection{Computational basis states}

\begin{figure}
 \centering
 \includegraphics[scale=0.57]{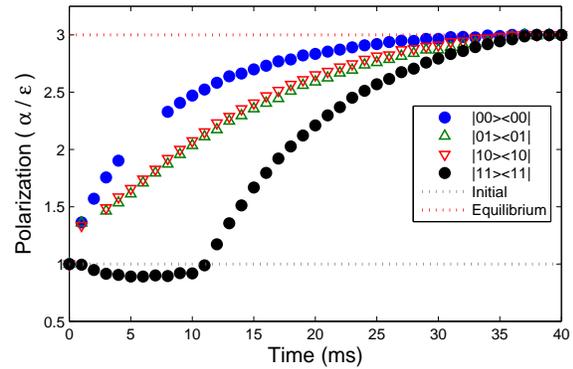}
 \caption{Polarization for computational basis states encoded in the spin $3/2$ nuclei.
 Also the initial and equilibrium limits to polarization are shown. Notice that initial polarizations are $\epsilon$,
 whereas final ones are $2I\epsilon$. Experimental data taken from previous experiments \cite{auccaise2008}. }
 \label{figpolarbc}
\end{figure}

In Fig. \ref{figpolarbc} we show some $\alpha$-polarization curves
for different initial states, namely the computational basis states
$|00\rangle$, $|01\rangle$, $|10\rangle$ and $|11\rangle$, obtained
from experimental relaxation results \cite{auccaise2008}.

The density matrix for a computational basis state is diagonal and
the ``polarization'' for each of these states is different. This
happens because the relaxation acts differently on each diagonal
elements of the density matrix. Since each state population relax at
a different rate \cite{sarthour2003}, the normalization will also be
different.

\subsection{Pseudo-entangled state}

Entanglement is a quantum correlation that reveals nonlocal properties of a given system. In order to compute such correlation or, in other words, to measure the degree of entanglement within the system, it is possible to use the concept of concurrence \cite{wootters98}. For a given density matrix $\rho$, the concurrence can be written as:

\begin{equation}\label{Conc}
\mathcal{C}(\rho) = \mbox{max}\left\{0, \sqrt{\lambda_{1}}-\sqrt{\lambda_{2}}-\sqrt{\lambda_{3}}-\sqrt{\lambda_{4}}\right\},
\end{equation}
where the $\lambda_{i}$s are the eigenvalues, in decreasing order, of
the matrix $R \equiv \rho(\sigma_{y}\otimes\sigma_{y})\rho^{*}(\sigma_{y}\otimes\sigma_{y})$. It is important to note that for density matrices like
\begin{equation}\label{EstadoX}
\rho = \left[ \begin{array}{cccc}
a & 0 & 0 & w \\
0 & b & z & 0 \\
0 & z^{*} & c & 0 \\
w^{*} & 0 & 0 & d \end{array} \right],
\end{equation}
named $X-$state, the concurrence has a analytical solution given by:
\begin{equation}\label{ConcX}
\mathcal{C}(\rho) = 2\,\mbox{max}\left\{0, \mathcal{C}^{I};
\mathcal{C}^{II}\right\},
\end{equation}
where $\mathcal{C}^{I} = |z|- \sqrt{a\,d}$ e $\mathcal{C}^{II} =
|w|- \sqrt{b\,c}$ \cite{fubini06,yu07}. 

For a NMR system, described by Eq.(\ref{eqpps}), it was shown that even being $|\psi\rangle$ a pure entangled state, i.e., a Bell basis state, the total density matrix is always separable \cite{braunstein99}. It happens because the polarization is very small ($\epsilon \approx 10^{-5}$, as said before), meaning that the density matrix is highly mixed. However, NMR is still capable to reproduce the correct dynamics of such pseudo-entangled states, since they behave like pure entangled ones \cite{linden01}.

Many works have been done about the dynamics of entanglement \cite{yu2002,simon2002,dur2004,yu09}. It was always expected that the dynamics of the concurrence would give an asymptotic decay of the degree of entanglement, since it is the typical behavior of the decoherence processes. However, for some quantum states it was reported a curious phenomenon, named sudden death of entanglement, where the degree of entanglement vanishes at a finite time \cite{yu09}. It has been recently measured in optical \cite{salles08} and atomic \cite{laurat07} systems. 

Fig. 4 shows a NMR experiment of the sudden death of entanglement of the pseudo-entangled state $|\psi^{+}\rangle=(|00\rangle + |11\rangle)/\sqrt{2}$, where its concurrence for such state is given by $\mathcal{C}^{I}$ term, using the normalization procedure described in Sec. III on the relaxation of $^{23}$Na. The continuous line is an exponential fit, included for comparison (asymptotic decay). We can clearly see that the concurrence vanishes at a finite time just below 20 ms, which is the same order of the transverse relaxation time of the sample.

\subsection{Discrete Wigner Function}

Another usual description to quantum states is the Wigner function
formalism, which consists on the representation of a quantum state
in the phase-space. Wigner functions can be obtained from the so called
``phase space point operators'' \cite{wootters87}. They can be defined
as the following expectation value:

\begin{equation} \label{eqdwf}
W(q,p)=\mbox{Tr}\left[\hat{\rho} \hat{A}(q,p)\right],
\end{equation}

\noindent where the $\hat{A}(q,p)$ are the (Hermitian) phase space
point operators and $\hat{\rho}$ is the density operator. Because we
are working with discrete spin systems, it is more adequate the use
of \emph{discrete Wigner functions} \cite{miquel02}. Such a
description is made upon periodic boundary conditions for both,
position and momentum (geometry of a torus on the phase-space). In
what follows, we will use the density operators (Eq.\ref{eqrhom})
relative to the relaxing pseudo-entangled state
$2^{-1/2}\left(|00\rangle+|11\rangle\right)$. The corresponding
phase-space point operators in matrix representation are shown on
Appendix \ref{apsop}.

The decoherence of superimposed localized wave packets is a well-known
subject in the literature \cite{zurek2002,schlosshauer07}. The Wigner function for
the cat-state exhibits two well separated positive peaks and a
non-classical interference pattern between them, which oscillates
between positive and negative values. Such oscillations vanish due
to the coupling of the system to the environment, leading to a
quantum-to-classical transition. Fig. \ref{figwig} shows NMR results
for the Wigner function of the cat-state
$2^{-1/2}\left(|00\rangle+|11\rangle\right)$ undergoing relaxation
towards equilibrium. The duration of the experiment is about 70 ms.
Fig. \ref{figwig}(a) shows the discrete phase-space representation of
this state (left), as well as the equilibrium state (right). Notice
that $W(0,p)$ and $W(6,p)$ are positive, whereas $W(7,p)$ is an
oscillating stripe between them. The oscillation of $W(2,p)$ and
$W(4,p)$ are due to the use of periodic boundary conditions and the
application of the discrete Fourier transform to obtain the momentum
\cite{miquel02}. Such ``images'' are important for the calculation
of the marginal momentum from $\sum_{q=0}^{2N-1}W(q,p)$ to guarantee
it as a regular probability distribution. Fig. \ref{figwig}(b) shows
the same plot as (a) for 3 and 12 ms, but now normalized according
to the method described in Sec. III. We can clearly see the changes
in the Wigner function caused by quadrupolar relaxation, in
particular the interference pattern.

Time-evolution of the momenta distribution can be seen on the bottom
of Fig. \ref{figwig}(c). The initial distribution are the red bullets,
({\color{red}$\bullet$}), whereas the blue one ({\color{blue}$\bullet$})
 are the final values. Intermediate values are the open dots
({\color{red}{$\circ$}}). From the plot we can see that momenta
evolves from a localized $p=0$ value,  to become homogeneous (better
seen in the inset, only for even values of momentum). This is an
indication that there is no localization \cite{henry2006}, in the
sense that it occurs in classically chaotic quantum maps. Finally,
Fig. \ref{figwig}(d) exhibits the time-evolution of the interference
term, $W(7,p)$. One can clearly see the decoherence by comparing the
initially red points ({\color{red}$\bullet$}) to the final blue ones
({\color{blue}$\bullet$}).

%########################################################################
\section{Conclusion}
\label{seccon}

In this work a method for the normalization of the initially pure
part of a pseudo-pure NMR state undergoing relaxation was presented.
The method allows following directly the pure part of $\rho$,
whereas keeping it a true density matrix, which is interesting for
the the purpose of NMR quantum information processing experiments.
We exemplified the method studying the relaxation of initially
pseudo-entangled states in which the phenomenon of sudden-death is
observed, and the decoherence of Wigner functions in phase-space.
The method is particularly useful for relaxation studies in the
context of NMR quantum information processing, an area which still
lacks experimental data.

\begin{acknowledgements}
The authors acknowledge the financial support of the Brazilian
Science Foundations CAPES, CNPq and FAPESP. A.M.S. would like to thanks the Government of Ontario-Canada. DOSP acknowledges FAPESP for financial support.
We also thank the support of the Brazilian network project National Institute for
Quantum Information.
\end{acknowledgements}

%################################################################################
\appendix
\section{Redfield theory of pure quadrupolar relaxation spin $3/2$}
\label{aRed}

The spin $3/2$ system dynamics under the presence of relaxation
processes can be described using the Redfield formalism for the
deviation density matrix \cite{abragam}. The Hamiltonian that
describes a $I > 1/2$ spin system in the laboratory frame,
considering Zeeman and first order quadrupolar interaction can be
expressed as

\begin{equation}
\mathcal{H} = - \hbar\omega_L I_z +
\frac{\hbar\omega_Q}{6}\left(3I_z^2 - I(I+1)\right).
\end{equation}

\noindent The first term describes the Zeeman interaction (with
Larmor frequency of $\omega_L$) and the second the static first
order quadrupolar interaction (with quadrupolar frequency of
$\omega_Q$). Hence, for nuclear spin $I=3/2$ the eigenvectors of the
Zeeman plus quadrupolar terms are $|3/2\rangle$, $|1/2\rangle$,
$|-1/2\rangle$ and $|-3/2\rangle$, which can be labeled as
$|00\rangle$, $|01\rangle$, $|10\rangle$ and $|11\rangle$,
corresponding to a two-qubit system. Here to simplify the
expressions of the solutions ($\sigma$) of the Redfield theory found in
\cite{auccaise2008} we utilize $\Delta=\sigma-\sigma_{\infty}$. The characterization of
the relaxation process is performed by means of reduced spectral
densities, which contain the information about the local field
fluctuations \cite{johan91,auccaise2008}. Furthermore, when
$|\omega_Q| \ll |\omega_L|$, the relaxation can be described by
three reduced densities at the Larmor frequency.  Below we show
$\Delta$ matrix elements in time function,

\begin{equation}
  \Delta_{12}\left(t\right)=\Delta_{12} e^{-t/\tau_{12}},
\label{eqRedf1}
\end{equation}

\begin{equation}
  \Delta_{03}\left(t\right)=\Delta_{03} e^{-t/\tau_{12}},
\label{eqRedf2}
\end{equation}

\begin{equation}
 \Delta_{01}\left(t\right)=\left[\frac{\left(1+e^{-t/\tau_2}\right)}{2} \Delta_{01} + \frac{\left(1-e^{-t/\tau_2}\right)}{2}\Delta_{23} \right]e^{-t/\tau_{01}},
\label{eqRedf3}
\end{equation}

\begin{equation}
 \Delta_{23}\left(t\right)=\left[\frac{\left(1-e^{-t/\tau_2}\right)}{2}\Delta_{01} + \frac{\left(1+e^{-t/\tau_2}\right)}{2}\Delta_{23} \right]e^{-t/\tau_{01}},
\label{eqRedf4}
\end{equation}

\begin{equation}
 \Delta_{02}\left(t\right)=\left[\frac{\left(1+e^{-t/\tau_1}\right)}{2}\Delta_{02} + \frac{\left(1-e^{-t/\tau_1}\right)}{2}\Delta_{13} \right]e^{-t/\tau_{02}},
\label{eqRedf5}
\end{equation}

\begin{equation}
 \Delta_{13}\left(t\right)=\left[\frac{\left(1-e^{-t/\tau_1}\right)}{2}\Delta_{02} + \frac{\left(1+e^{-t/\tau_1}\right)}{2}\Delta_{13} \right]e^{-t/\tau_{02}},
\label{eqRedf6}
\end{equation}

\begin{eqnarray}
  \Delta_{00}\left(t\right)=& \left[\frac{\left( e^{-t/\tau_{2}} +e^{-t/\tau_{1}}\right)}{2}\Delta_{00}+
                                   \frac{\left(-e^{-2t/\tau_{12}}+e^{-t/\tau_{2}}\right)}{2}\Delta_{11}+\right. \nonumber \\
                           &+\left.\frac{\left(-e^{-2t/\tau_{12}}+e^{-t/\tau_{1}}\right)}{2}\Delta_{22} \right],
\label{eqRedf7}
\end{eqnarray}

\begin{eqnarray}
  \Delta_{11}\left(t\right)=& \left[\frac{\left( e^{-t/\tau_{2}} -e^{-t/\tau_{1}}\right)}{2}\Delta_{00}+
                                   \frac{\left( e^{-2t/\tau_{12}}+e^{-t/\tau_{2}}\right)}{2}\Delta_{11}+\right. \nonumber \\
                           &+\left.\frac{\left( e^{-2t/\tau_{12}}-e^{-t/\tau_{1}}\right)}{2}\Delta_{22} \right],
\label{eqRedf8}
\end{eqnarray}

\begin{eqnarray}
  \Delta_{22}\left(t\right)=& \left[\frac{\left(-e^{-t/\tau_{2}} +e^{-t/\tau_{1}}\right)}{2}\Delta_{00}+
                                   \frac{\left( e^{-2t/\tau_{12}}-e^{-t/\tau_{2}}\right)}{2}\Delta_{11}+\right. \nonumber \\
                           &+\left.\frac{\left( e^{-2t/\tau_{12}}+e^{-t/\tau_{1}}\right)}{2}\Delta_{22} \right],
\label{eqRedf9}
\end{eqnarray}

\begin{eqnarray}
  \Delta_{33}\left(t\right)=&-\left[\frac{\left( e^{-t/\tau_{2}} +e^{-t/\tau_{1}}\right)}{2}\Delta_{00}+
                                   \frac{\left( e^{-2t/\tau_{12}}+e^{-t/\tau_{2}}\right)}{2}\Delta_{11}+\right. \nonumber \\
                           &+\left.\frac{\left( e^{-2t/\tau_{12}}+e^{-t/\tau_{1}}\right)}{2}\Delta_{22} \right],
\label{eqRedf10}
\end{eqnarray}

\noindent where the elements $\Delta_{kl}$ from right side are
initial values of them ($\Delta=\sigma_0 -\sigma_{\infty}$).
 In addition, the characteristic times $\tau$
are defined as function of the inverse averaged reduced spectral
densities \cite{auccaise2008}. The characteristic times were determined from previous work as: $\tau_{01}= 4.6$, $\tau_{02}=4.7$, $\tau_{12}=11.1$, $\tau_{12}=20.8$ and $\tau_2=23.8$ in miliseconds.

\section{Matrix representation to phase space point operators}
\label{apsop}

The phase space point operators are a set of Hermitian operators, which gives all marginal distributions obtained from the Wigner function \cite{miquel02}. They are defined on a phase space grid of $2N \times 2N$ points

\begin{equation}
\hat{A}(q,p)=\frac{1}{2N}\hat{U}^q\hat{R}\hat{V}^{-p}\mbox{exp}(i\pi qp/N),
\end{equation}

\noindent where $\hat{U}$ is a cyclic shift operator in the computational basis ($\hat{U}|q\rangle=|q+1\rangle$), $\hat{V}$ is a cyclic shift operator in the basis related to the computational basis via the discrete Fourier transform and $\hat{R}$ is the reflection operator ($\hat{R}|q\rangle=|N-q\rangle$). The phase space point operator form a complete orthonormal basis of the space of operators and

\begin{equation}
\hat{A}(q+\xi_qN,p+\xi_pN)=\hat{A}(q,p)(-1)^{\xi_pq+\xi_qp+\xi_q\xi_pN},
\end{equation}

\noindent for $\xi_q,\xi_q=0,1$. For this reason, it is clear that the $N^2$ phase space point operators corresponding to the first $N \times N$ subgrid of the determine the rest.

NMR $3/2$ spin system have four eigenstates (computational basis, $|3/2\rangle$, $|1/2\rangle$, $|-1/2\rangle$ and $|-3/2\rangle$). Thus, the Hilbert space dimension is $N=4$. The operators $\hat{A}(q,p)$ in matrix representation are shown below:

\begin{eqnarray}
A(0,p)=\frac{1}{8}
       \left(
        \begin{array}{cccc}
        1 & 0 & 0      & 0\\
        0 & 0 & 0      & i^p\\
        0 & 0 & (-1)^p & 0\\
        0 & (-i)^p & 0      & 0
        \end{array}
       \right),
\end{eqnarray}
\begin{eqnarray}
A(1,p)=\frac{e^{i\pi p/N}}{8}
       \left(
        \begin{array}{cccc}
        0 & 1 & 0     & 0\\
        i^p & 0 & 0  & 0\\
        0 & 0 & 0 & (-1)^p\\
        0 & 0 & (-i)^p    & 0
        \end{array}
       \right),
\end{eqnarray}
\begin{eqnarray}
A(2,p)=\frac{e^{i2\pi p/N}}{8}
       \left(
        \begin{array}{cccc}
        0 & 0 & 1      & 0\\
        0 & i^p & 0      & 0\\
        (-1)^p & 0 & 0 & 0\\
        0 & 0 & 0      & (-i)^p
        \end{array}
       \right),
\end{eqnarray}
\begin{eqnarray}
A(3,p)=\frac{e^{i3\pi p/N}}{8}
       \left(
        \begin{array}{cccc}
        0 & 0 & 0      & 1\\
        0 & 0 & i^p      & 0\\
        0 & (-1)^p & 0 & 0\\
        (-i)^p & 0 & 0      & 0
        \end{array}
       \right).
\end{eqnarray}

\noindent  The projection of the density matrix $\rho$ in these matrices (Eq.\ref{eqdwf}) yields a grid with only real values which represents the density matrix in discrete phase space.

% #################################################################################


\begin{thebibliography}{}
\bibitem{pazzurek01}
     Paz, J.P., Zurek, W.H.: Environment-induced and the transition from quantum to classical. In: Kaiser, R., Westbrook, C., Davids, F. (eds.) Coherent Matter Waves, Proceedings of the Les Houches Session LXXII, pp. 533-614. Springer Verlag, Berlin (2001)

\bibitem{nielsen2000}  %4
     Nielsen, M.A., Chuang, I.L.: Quantum Computation and Quantum Information. Cambridge University Press, Cambridge (2000)

\bibitem{vandersypen04}
     Vandersypen, L.M.K, Chuang, I.L.: NMR techniques for quantum control and computation. Rev. Mod. Phys. \textbf{76}, 1037-1069 (2004)

\bibitem{zurek82}
     Zurek, W.H.: Environment-induced superselection rules. Phys. Rev. D \textbf{26}, 1862-1880 (1982)

\bibitem{teklemariam03}
     Teklemariam, G., Fortunato, E.M., L\'opez, C.C., Emerson, J., Paz, J.P., Havel, T.F., and Cory, D.G.: Method for modeling decoherence on a quantum-information processor. Phys. Rev. A \textbf{67}, 062316 (2003)

\bibitem{nagele08}
     N\"agele, P., Campagnano, G., Weiss, U.: Dynamics of dissipative coupled spins: decoherence, relaxation and effects of a spin-boson bath. New J. Phys. \textbf{10}, 115010 (2008)

\bibitem{ivan2007}
    Oliveira, I.S., Bonagamba, T.J., Sarthour, R.S., Freitas, J.C.C., deAzevedo, E.R.: NMR quantum information processing. Elsevier, Amsterdam (2007)

\bibitem{braunstein99}
	Braunstein, S.L., Caves, C.M., Jozsa, R., Linden, N., Popescu, S., Schack, R.: Separability of Very Noisy Mixed States and Implications for NMR Quantum Computing. Phys. Rev. Lett. \textbf{83}, 1054 (1999)

\bibitem{suter2009}
	Suter, D., Mahesh, T.S.: Spins as qubits: Quantum information processing by nuclear magnetic resonance. J. Chem. Phys. \textbf{128}, 052206 (2008)

\bibitem{cory97}
    Cory, D.G., Fahmy, A.F., Havel, T.F.: Ensemble quantum computing by NMR spectroscopy. Proc. Natl. Acad. Sci. USA, \textbf{94}, pp. 1634-1639 (1997)

\bibitem{sarthour2003}
    Sarthour, R.S., deAzevedo, E.R., Bonk, F.A., Vidoto, E.L.G., Bonagamba, T.J., Guimarães, A.P.,  Freitas, J.C.C., Oliveira, I.S.: Relaxation of coherent states in a two-qubit NMR quadrupole system. Phys. Rev. A \textbf{68}, 022311 (2003)

\bibitem{ghosh2005}
    Ghosh, A., Kumar, A.: Relaxation of pseudo pure states: the role of cross-correlations. J. Magn. Res. \textbf{173}, 125-133 (2005)

\bibitem{boulant03}
    Boulant, N., Havel, T.F., Pravia, M.A., Cory, D.G.: Robust method for estimating the Lindblad operators of a dissipative quantum process from measurements of the density operator at multiple time points. Phys. Rev. A \textbf{67}, 042322 (2003)

% \bibitem{souza09}
%     A.M. Souza, R. Auccaise, A. Gavini-Viana, J. Teles, E.R. deAzevedo, T.J. Bonagamba, I.S. Oliveira and R. S. Sarthour,
%     \emph{submitted}.

\bibitem{garcia-mata2005}
    García-Mata, I., Saraceno, M., Spina, M.E., Carlo, G.: Phase-space contraction and quantum operations. Phys. Rev. A \textbf{72}, 062315 (2005)

\bibitem{leung}
    Leung, D., PhD. theses, Stanford University (2002)

\bibitem{fortunato2002}
    Fortunato, E.M., Pravia, M.A., Boulant, N., Teklemarian, G., Havel, T.F., Cory, D.G.: Design of strongly modulating pulses to implement precise effective Hamiltonians for quantum information processing. J. Chem. Phys. \textbf{116}, 7599 (2002)

\bibitem{teles2007}  %31
     Teles, J., deAzevedo, E.R., Auccaise, R., Sarthour, R.S., Oliveira, I.S., Bonagamba, T.J.: Quantum state tomography for quadrupolar nuclei using global rotations of the spin system. J. Chem. Phys. \textbf{126}, 154506 (2007)

\bibitem{auccaise2008}  %31
     Auccaise, R., Teles, J., Sarthour, R.S., Bonagamba, T.J., Oliveira, I.S., deAzevedo, E.R.: A study of the relaxation dynamics in a quadrupolar NMR system using Quantum State Tomography. J. Magn. Res. \textbf{192}, 17-26 (2008)

\bibitem{johan91}
     van der Maarel, J.R.C.: Relaxation of spin quantum number S=3/2 under multiple-pulse quadrupolar echoes. J. Chem. Phys. \textbf{94}, 4765-4775 (1991)

\bibitem{wootters98}
     Wootters, W.K.: Entanglement of formation of an arbitrary state of two qubits. Phys. Rev. Lett. \textbf{80}, 2245-2248 (1998)    
     
\bibitem{fubini06}
	Fubini, A., Roscilde, T., Tognetti, V., Tusa, M., Verrucchi, P.: Reading entanglement in terms of spin configurations in quantum magnets. Eur. Phys. J. D \textbf{38}, 563-570 (2006)

\bibitem{yu07}
	Yu, T., Eberly, J. H.: Evolution from entanglement to decoherence. Quantum Inf. Comput. \textbf{7}, 459 (2007)

\bibitem{linden01}
	Linden, N., Popescu, S.: Good Dynamics versus Bad Kinematics: Is Entanglement Needed for Quantum Computation?. Phys. Rev. Lett. \textbf{87}, 047901 (2001)
	
\bibitem{yu2002}
     Yu, T., Eberly, J.H.: Phonon decoherence of quantum entanglement: Robust and fragile states. Phys. Rev. B \textbf{66}, 193306 (2002)

\bibitem{simon2002}
     Simon, C., Kempe, J.: Robustness of multiparty entanglement. Phys. Rev. A \textbf{65}, 052327 (2002)

\bibitem{dur2004}
     D\"ur, W., Briegel, H.J.: Stability of Macroscopic Entanglement under Decoherence. Phys. Rev. Lett. \textbf{92}, 180403 (2004)
     
\bibitem{yu09}
	 Yu, T., Eberly, J. H.: Sudden Death of Entanglement. Science \textbf{323}, 598 - 601 (2009)
     
\bibitem{salles08}
     Salles, A., de Melo, F., Almeida, M.P., Hor-Meyll, M., Walborn, S.P., Souto Ribeiro, P.H., Davidovich,  L.: Experimental investigation of the dynamics of entanglement: Sudden death, complementarity, and continuous monitoring of the environment. Phys. Rev. A \textbf{78}, 022322 (2008)
     
\bibitem{laurat07}
	 Laurat, J., Choi, K.S., Deng, H., Chou, C.W., Kimble, H.J.: Heralded Entanglement between Atomic Ensembles: Preparation, Decoherence, and Scaling. Phys. Rev. Lett. \textbf{99}, 180504 (2007)

\bibitem{wootters87}
     Wootters, W.: A Wigner-function formulation of finite-state quantum mechanics. Ann. of Phys. (N.Y.) \textbf{176}, 1-21 (1987)
     
\bibitem{miquel02}
     Miquel, C., Paz, J.P., Saraceno, M.: Quantum computers in phase space. Phys. Rev. A, \textbf{65}, 062309 (2002)

\bibitem{zurek2002}
     Zurek, W.H.: Decoherence and the transition from quantum to classical-revisited. \textit{Los Alamos Science}, \textbf{27}, 86 (2002) [quant-ph/0306072]
     
\bibitem{schlosshauer07}
	 Schlosshauer, M.: Decoherence and the quantum-to-classical transition. Springer-Verlag, Heidelberg (2007)

\bibitem{henry2006}
     Henry, M.K., Emerson, J., Martinez, R., Cory, D.G.: Localization in the quantum sawtooth map emulated on a quantum-information processor. Phys. Rev. A \textbf{74}, 062317 (2006)

\bibitem{abragam}
    Abragam, A.: The principles of nuclear magnetism, Clarendon Press, Oxford (1994)

\end{thebibliography}


\begin{figure}
 \centering
 \includegraphics[scale=0.56]{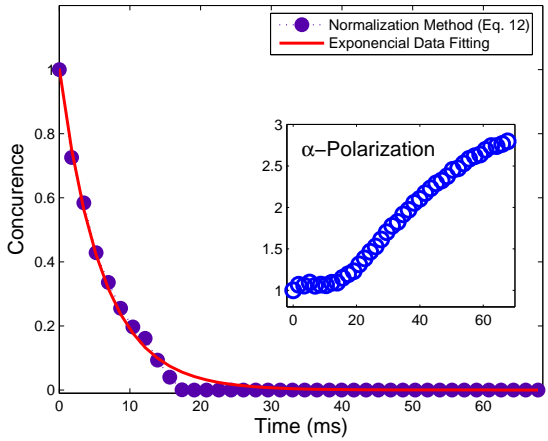}
 \caption{Concurence decay of pseudo-entangled initial state
 $2^{-1/2}\left(|00\rangle+|11\rangle\right)$.
 The data were obtained from experimental deviation density matrices of the relaxation of
 $^{23}$Na nuclei, normalized by
 the procedure developed in the section \ref{secnormmet}. The solid line is the exponential fitting, included for comparison. The $\alpha$-polarization curve utilized to normalize the relaxing pseudo-cat state are the inside plot.}
 \label{figconc}
\end{figure}

\begin{figure}
\centering
    \mbox{
    \subfigure[Theoretical: initial and equilibrium]{\label{figtheo}
      \centering \includegraphics[scale=0.3]{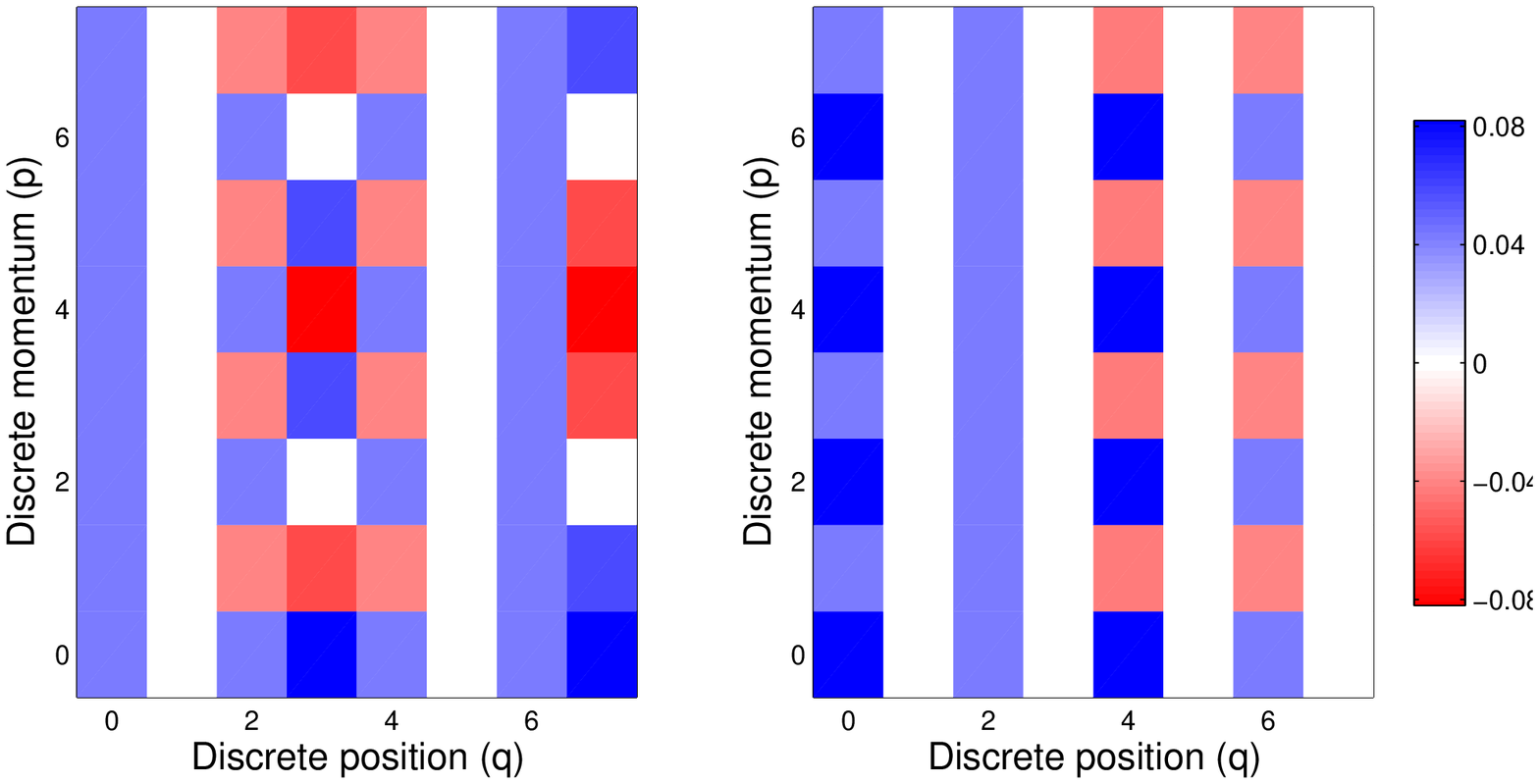}} %\qquad
    }
	\mbox{
    \subfigure[Experimental: $3$ ms and $12$ ms]{\label{figexps}
      \centering \includegraphics[scale=0.3]{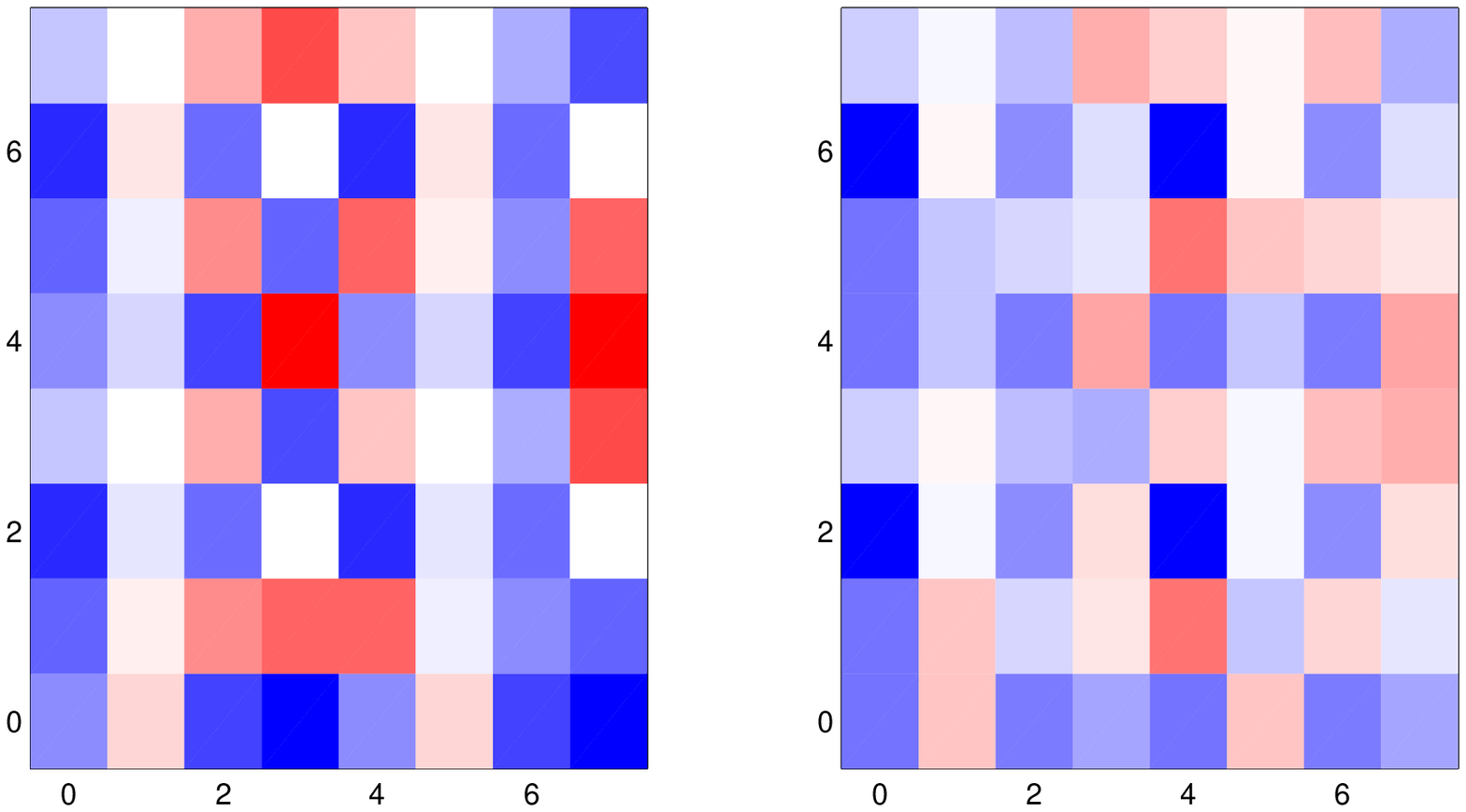}} %\qquad height=0.18\textheight,width=0.4\textwidth
    }
    \mbox{
    \subfigure[Momentum distribution ]{\label{figprob}
      \centering \includegraphics[scale=0.45]{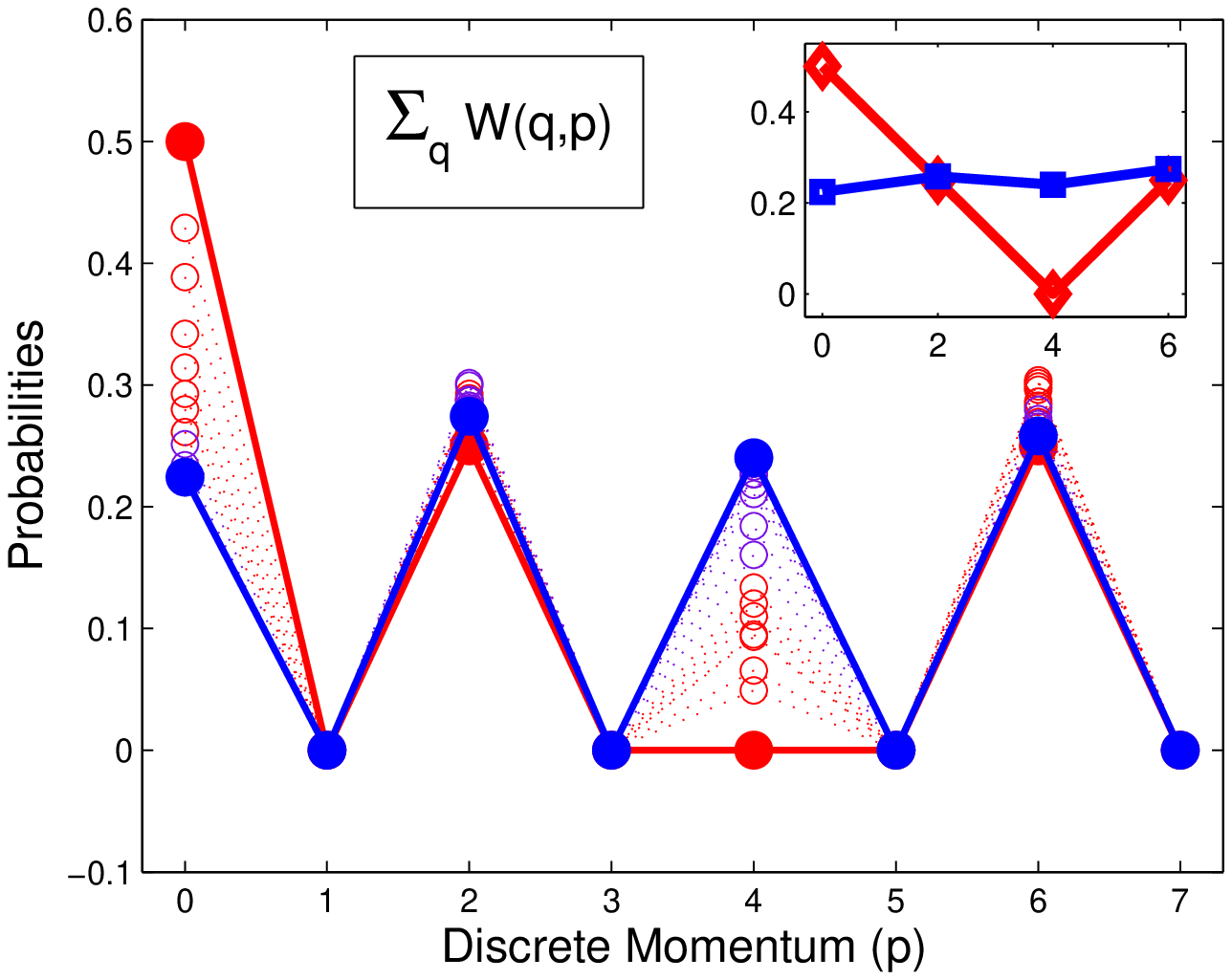}} %\qquad
    } \\
	\mbox{
    \subfigure[The oscillating term of the Wigner function]{\label{w11}
      \centering \includegraphics[scale=0.45]{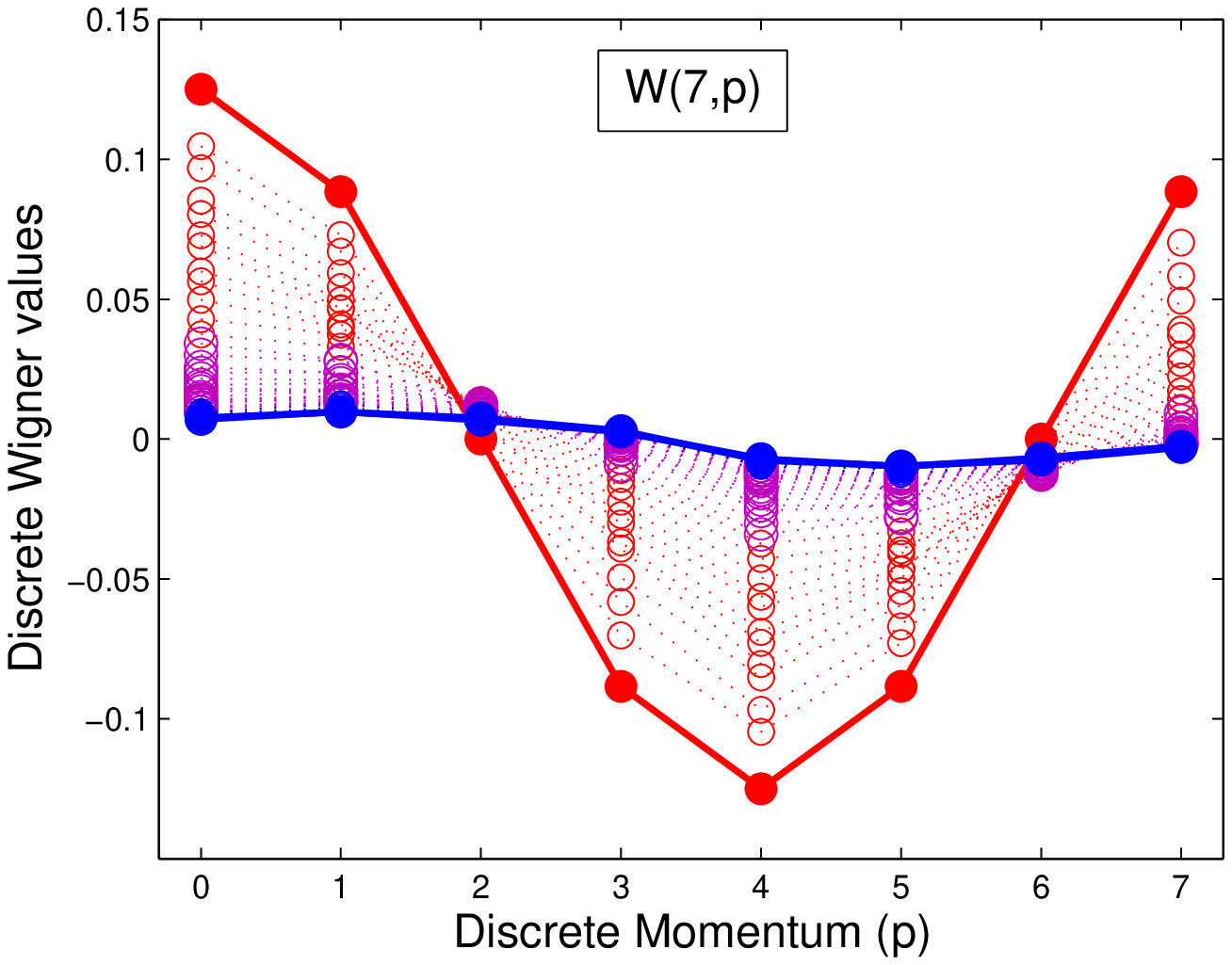}}
    }
  \caption{The decoherence of the pseudo-entangled state observed from quadrupolar relaxation.
  In (a) and (b) are theoretical and experimental discrete Wigner functions, where the blue(red)
  squares represent the positive(negative) values and white squares are null values. Horizontal(vertical) axis is the position(momentum) coordinates.
  The temporal evolution are shown in (c) for the discrete momentum distribution and in (d)
  for the oscillating term $W(7,p)$. Red(Blue) dots are the initial(final) values and the inside plot are the initial(red) and final(blue)
  momentum distribution for only even momentum.}
\label{figwig}
\end{figure}

\end{document}